\documentclass[epj]{svjour}
\usepackage{graphics}
\begin{document}
\newcommand{\mathbold}[1]{\mbox{\boldmath $\bf#1$}}
\title{Including virtual photons in strong interactions}
\author{A.~Rusetsky\inst{1,2}
}                     
%
%
\institute{Institute for Theoretical Physics, University of Bern,
Sidlerstrasse 5, 3012 Bern, Switzerland
\and 
High Energy Physics Institute,
Tbilisi State University, University St.~9, 380086 Tbilisi, Georgia}
\date{Received: date / Revised version: \today}
%
\abstract{
In the perturbative field-theoretical models
we investigate the inclusion of the electromagnetic
interactions into the purely strong theory that describes hadronic
processes. In particular, we study the
convention for splitting electromagnetic and strong interactions and the
ambiguity of such a splitting. 
The issue of the interpretation of the parameters
of the low-energy effective field theory in the presence of electromagnetic
interactions is addressed, as well as the scale and gauge dependence of the
effective theory couplings. We hope, that the results of these studies
are relevant for the electromagnetic sector of ChPT.
} 
\maketitle
\section{Introduction}
\label{intro}

The systematic approach to take into account the electromagnetic
corrections in the low-energy processes involving hadrons is based on the
Chiral Perturbation Theory (ChPT) with virtual photons~\cite{Urech}, which is the
low-energy effective theory of the Standard Model in the hadron
sector. Despite of been widely used in the applications,
the procedure of the construction of the
effective chiral Lagrangian with virtual photons from QCD is not free from
conceptual difficulties. In particular, we mention the following points:

\begin{itemize}

\item[i)]
The effective Lagrangian with virtual photons operates with quantities
like the quark masses $m_q$, the parameter $B$ which is related with
the quark condensate, the pion decay constant in the chiral limit $F$,
etc. It is usually not specified, to which underlying theory --
QCD+photons, or pure QCD -- do these quantities refer. Obviously, if
one interprets $m_q$ to be the running quark masses in the full theory
(including virtual photons), then the quark mass ratio $m_u/m_d$ in ChPT
Lagrangian turns out to be 
QCD-scale dependent. One encounters similar problem in relating $B$ to
the quark condensate, since $\langle 0|\bar u u|0\rangle$ and
$\langle 0|\bar d d|0\rangle$ run differently for $e\neq 0$.

\item[ii)]
According to the commonly used terminology, in the effective theory 
the combined contribution of the explicit photon loops and the
electromagnetic effective couplings is called ``electromagnetic corrections'', and
the rest is referred to as ``strong piece''. This implies, that 
exactly the latter survives when electromagnetic interactions are
switched off. The problem is, however, how to rigorously define the
theory with electromagnetic interactions switched off, even for the
underlying QCD. To be precise, note that the initial theory with
photons had four parameters: strong coupling constant $g$, fine
structure constant $\alpha$, and the quark masses $m_u$ and $m_d$.
The theory without photons has only three parameters: the strong
coupling constant $\bar g$ which is obtained from $g$  in the
limit $\alpha\to 0$, and the quark masses $\bar m_u$, $\bar m_d$ to which
$m_u$ and $m_d$ converge in the same limit. The question is, how to
determine $\bar g$, $\bar m_u$ and $\bar m_d$ 
which define the theory in the limit
$\alpha\to 0$? And, once this is done, how the above definition
translates into the commonly used splitting of the low-energy
effective theory?

\item[iii)]
Although it has been pointed out~\cite{BP,Moussallam}, that some of the
electromagnetic effective couplings do indeed contain scale- and
gauge-dependence which is determined by the underlying QCD, a systematic
study of this phenomenon is still lacking.
\end{itemize}

\vspace*{-.1cm}

Although some of the issues mentioned here, have been already addressed
and/or mentioned in the literature (see, e.g.~\cite{BP,Moussallam}), 
a focused discussion of the conceptual problems in ChPT with virtual photons,
to the best of our knowledge, does not exist so far.
The aim of our investigations, which are very briefly surveyed in this work,
is to set up notions for creation of 
such a coherent framework. We would like to stress in addition, 
that such investigations are an important ingredient in carrying out the
consistent calculation of isospin-breaking corrections
in the context of hadronic atom problem~\cite{Bern}.
In order to demonstrate the general procedure,
we do not consider QCD here -- rather, we restrict ourselves to the
simple perturbative models, where one may include
electromagnetic interactions in a crystal clear manner. We believe,
that the lessons one learns from these models, provide the necessary clue in a
much more complicated case of QCD.
A detailed discussion of the issues raised in this work, is given in the
forthcoming publication~\cite{Scimemi}.

\section{Convention for the splitting}
\label{sec:1}

In this section, the following question is addressed: suppose, that 
one has the field theory which describes both electromagnetic and
non-electromagnetic (referred hereafter as to the ``strong'') interactions.
How does one systematically split strong and electromagnetic contributions in
the quantities which are calculated within this theory? In order to answer
this question, we find it useful to consider a simple perturbative model - 
the Yukawa model. The splitting convention which is explained in this model,
is general and can be applied in the context of any other field-theoretical model.  

The Yukawa model considered in this work,
describes the doublet of ``colored'' fermions 
$\bar\Psi=(\bar u^i,\bar d^i)$, where the ``color'' index $i=1,2$.
The fermions interact with the triplet of boson fields through the usual
Yukawa coupling
${\cal L}_{\rm str}=g\bar\Psi{\mathbold\tau}\mathbold{\phi}\Psi$, which is
characterized by a ``strong'' constant $g$. Further, the
fermions whose charge matrix is given by $eQ\doteq e\,{\rm diag}(Q_u,Q_d)$,
interact with the photon field in a standard manner, whereas the bosons
described by the field $\mathbold{\phi}$, are neutral.
For the renormalization, $\overline{\rm MS}$ scheme is used.

We demonstrate the idea of the splitting on the example
of the physical mass of the fermion fields, 
given by the position of the pole in the propagator. Below, 
we restrict ourselves to the one-loop order.
Denoting the masses by $M_q$ where $q=u,d$, we find
\begin{eqnarray}\label{mphys}
M_q&=&m_q
\bigl[1+\frac{3}{16\pi^2}(3g_r^2-2e_r^2Q_q^2)\ln\frac{m_q}{\mu}
+A_1g_r^2
\nonumber\\
&+&A_2Q_q^2e_r^2\bigr]+O(g_r^4,e_r^2g_r^2,e_r^4)\, ,
\end{eqnarray}
where $g_r,~e_r$ denote the renormalized couplings,
$\mu$ is the scale of the dimensional
regularization, $m_q$ stands for the running fermion mass. 
Further, $A_1$ is a known function of the boson-fermion mass ratio which is
independent of the scale $\mu$ at this order (the
explicit expression for this quantity is not needed), and 
$A_2=(16\pi^2)^{-1}$.

The physical masses become scale independent, provided that the 
masses $m_q$ run properly with the scale,
\begin{eqnarray}\label{eq:runm}
\mu\frac{dm_q}{d\mu}=\frac{3}{16\pi^2}(3g_r^2-2e_r^2Q_q^2)m_q
+O(g_r^4,e_r^2g_r^2,e_r^4)\, .
\end{eqnarray}
The couplings $g_r,e_r$ are scale independent at this order in the 
perturbative expansion.
Once the running mass $m_q$ is known at a some scale, the physical mass
 $M_q$ is fixed in terms of the coupling constants $g_r,e_r$ and of the
 running boson and fermion masses at this order in the perturbative expansion.

We now discuss the splitting of the physical masses into a strong and 
an electromagnetic part. This splitting should divide the mass into 
a piece that one would calculate in a theory with 
no electromagnetic interactions, and a part proportional to $e_r^2$:
$M_q=\bar{M}_q+e_r^2M_q^1+O(e_r^4)$.
Here and below, barred quantities refer to the theory at $e_r=0$.
The first term on the right hand side is 
\begin{eqnarray}
\bar{M}_q=\bar{m}_q\big[1+\frac{9\bar{g}_r^2}{16\pi^2}\ln\frac{\bar{m_q}}
{\mu}+A_1\bar g_r^2\big]+O(\bar{g}_r^4)\, .
\end{eqnarray}
This part is scale independent by itself, 
provided that  the mass $\bar{m}_q$ runs according to renormalization group
equation~(\ref{eq:runm}) for $e_r=0$.
On the other hand,  
$\bar{g}_r$ is scale independent in this approximation. 
We see,
that one has to fix a boundary condition 
in order to determine unambiguously $\bar{M}_q$. As a natural condition, 
we choose the running mass $\bar{m}_q$ to coincide with the running 
mass $m_q$ in the full theory at a some scale:
$m_q(\mu)=\bar m_q(\mu;\mu_1)~{\rm at}~\mu=\mu_1$ (here, we have explicitly
indicated the $\mu_1$-dependence of the barred quantities).
%
%
With the use of the above matching condition, we 
express $m_q(\mu)$ through $\bar m_q(\mu;\mu_1)$, 
and insert the result into the expression for the mass $M_q$. 
Identifying ${g}_r$ with $\bar{g}_r$ at this order,  we find that
\begin{eqnarray}\label{split-phys}
\bar{M}_q&=&\bar{m}_q(\mu;\mu_1)\big[1+\frac{9\bar{g}_r^2}{16\pi^2}
\ln\frac{\bar{m}_q}{\mu}+\bar g_r^2A_1\big]+O(\bar{g}_r^4)\, ,\nonumber\\
M_q^1&=&-\bar{m}_q(\mu;\mu_1)\big[\frac{6}{16\pi^2}\ln\frac{\bar{m}_q}
{\mu_1}-A_2\big]Q_q^2+O(\bar g_r^2)\, .
\end{eqnarray}
This splitting has the desired properties: Each term on the 
right-hand side is scale-independent. However, as is explicitly seen 
in the contribution proportional to $e_r^2$, the splitting does depend on the 
{\em matching scale} $\mu_1$. Indeed, one has
\begin{eqnarray}
\mu_1\frac{d\bar{M}_q}{d\mu_1}&=&
 -\mu_1\frac{d[e_r^2M_q^1]}{d\mu_1}
=-\frac{6e_r^2Q_q^2}{16\pi^2}\bar{M}_q\, .
\end{eqnarray}
In other words, both terms in the splitting depend on the scale $\mu_1$. 
This scale dependence is of order $e_r^2$ in the approximation considered.
The sum $M_q$ is of course independent of the matching scale.

A similar splitting may be considered for 
the running masses themselves. Indeed, expressing $m_q(\mu)$ through the
running mass in the purely strong theory $\bar m_q(\mu;\mu_1)$ gives
\begin{eqnarray}\label{eq:splittingmrun}
m_q(\mu)=\bar{m}_q(\mu;\mu_1)\big[1-\frac{6e_r^2Q_q^2}{16\pi^2}
\ln\frac{\mu}{\mu_1}\big]\, .
\end{eqnarray}
This result is the analogue of the relation (\ref{split-phys}) for the 
physical masses. 
It shows that the splitting of the running masses into a part that 
runs with the 
strong interaction alone, and a piece proportional to $e_r^2$, 
depends on the matching scale.

\begin{figure}
\vspace*{.4cm}
\resizebox{0.40\textwidth}{!}{%
  \includegraphics{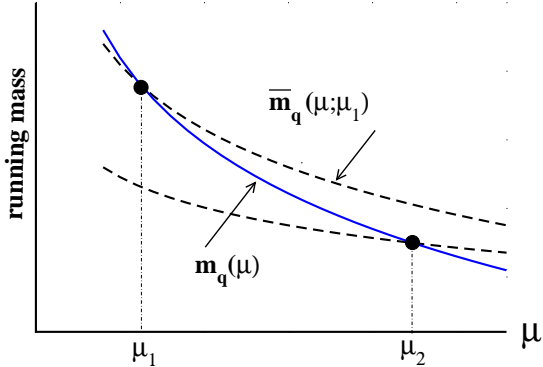}
}
\vspace*{.2cm}
\caption{The matching condition. The solid 
line represents the running of the mass $m_q$ in the full theory according 
to (\ref{eq:runm}), whereas the dashed lines display the running 
of $\bar m_q$.}
\label{fig:1}       
\end{figure}

The dependence of the splitting on the scale $\mu_1$ originates 
in the different running of the masses in the full theory and in 
the approximation when $e_r=0$. 
This is illustrated in Fig.~\ref{fig:1}.
The solid line refers to the running of the 
mass $m_q$ in the full theory, whereas the dashed lines represent 
the running of $\bar{m}_q$. Because, for a fixed value of the scale $\mu$, 
the running mass $\bar{m}_q$ depends on the matching scale chosen, 
the mass $\bar{M}_q$ does so as well.

One may wonder whether there is 
a way to split the pole mass in a unique manner. The reason why this 
is not the case is the following. In the Yukawa model considered here, 
the pole mass is proportional to $m_q$, which itself depends on the 
scale $\mu$. In order to compare this mass with the corresponding quantity 
at $e_r=0$, one has to compare two quantities that run differently, 
$\bar m_q$ and $m_q$. This running is itself a one-loop effect. There is 
therefore no possibility to avoid the ambiguity. 

Finally, we mention that
the splitting of the parameters of the theory (masses and couplings)
along the lines demonstrated above 
can be performed from a knowledge of the 
relevant beta-functions of the masses and of the coupling constants
 to any order in the perturbative expansion~\cite{Scimemi}.

\section{Splitting in the effective theory}
\label{sec:2}

On the example of the linear $\sigma$-model (L$\sigma$M)
with electromagnetic interactions
we illustrate, how the prescription for splitting of electromagnetic
and strong interactions does translate to the language of the low-energy
effective theory.
The calculations are done at one loop.
In the limit of the large $\sigma$-mass, the model is equivalent to ChPT with
the particular values of the couplings, expressed explicitly 
through the parameters of
the initial model. This allows one to study the consequence of the splitting,
performed in L$\sigma$M, for the couplings of the effective theory.
The dependence of these couplings on the renormalization scale in L$\sigma$M,
as well as on the gauge parameter, can be also studied.
Below, we briefly list our conclusions obtained from these investigations.
Detailed discussion is given in Ref.~\cite{Scimemi}.

\begin{itemize}

\item[i)]
The splitting which is carried out in the underlying theory, is directly
translated to the level of the low-energy effective Lagrangian of this theory.
The $\mu_1$-ambiguity of the parameters (masses and coupling constants) of the
underlying {\it purely strong} theory which is matched to the theory with
virtual photons at a scale $\mu=\mu_1$, is lumped in the couplings of
the effective Lagrangian. These couplings have to be expressed in terms 
of the parameters of the purely strong theory
and, possibly, some additional parameters that have to be
introduced when the electromagnetic interactions are turned on. The advantage
of doing so is, that the different parts of the Lagrangian then exactly
describe the low-energy limit of the purely strong theory, and what is called
the electromagnetic corrections. The price for this is just the
above-mentioned $\mu_1$-dependence of the effective couplings.

\item[ii)]
When the electromagnetic interactions are turned on, some quantities like,
e.g. the matrix elements of the vector current, start to be scale- and
gauge-dependent. At the level of the effective theory, this dependence is
systematically transformed into the scale- and gauge-dependence of
the couplings of the Lagrangian.

\item[iii)]
Most of the above conclusions can be straightforwardly applied to QCD without
any change. The splitting of the quark masses, condensates, etc proceeds along
the lines similar to those described in section~\ref{sec:1}.
Further, the parameters of the low-energy effective Lagrangian of QCD in the
strong sector refer to the pure QCD rather then to QCD+photons: e.g. the quark
masses are the masses in pure QCD, etc.
The price to pay is, that the couplings in ChPT
depend on the matching scale $\mu_1$. The
($\mu_1$-dependent) results of calculations for any physical quantity,
based only on the strong part of
the effective Lagrangian, exactly reproduce the results that would 
be obtained for the same quantity in pure QCD matched to QCD+photons at
$\mu=\mu_1$.

\item[iv)]
It is important to note, that the $\mu_1$-dependence 
puts natural limitations on the accuracy at which the
couplings of the effective Lagrangian 
can be determined from the physical data which, of course, contains no
$\mu_1$ dependence. In principle, such a dependence should be observed if
the couplings are theoretically derived from the underlying QCD e.g.
via the lattice simulations.

\end{itemize}

{\em Acknowledgments.}
Present work is based on the results obtained in collaboration with
  J. Gasser, I. Mgeladze and I. Scimemi~\cite{Scimemi}. 
We have profited from
 unpublished notes by H.~Leutwyler, whom we thank in addition for
informative discussions on the subject. We are thankful to G. Ecker, 
H. Neufeld, B. Moussallam and U. Wiese for interesting discussions.
This work was supported in part by the Swiss National Science
Foundation, and by TMR, BBW-Contract No. 97.0131  and  EC-Contract
No. ERBFMRX-CT980169 (EU\-RO\-DA\-$\Phi$NE), and by SCOPES Project
No. 7UZPJ65677.

\end{document}